\documentclass[a4paper,12pt]{article}
\usepackage{cite,hyperref}
\usepackage{amsmath,amssymb,amsfonts}
\usepackage{epsfig}

\def\bsigma{\mbox{\protect\boldmath $\sigma$}}
\newcommand{\reff}[1]{(\ref{#1})}
\DeclareMathAlphabet{\mathbi}{OT1}{cmr}{bx}{it}
\newcommand{\pos}[1]{\mathbi{#1}}

\title{Shape dependence of the finite-size scaling limit 
in a strongly anisotropic $O(\infty)$ model}

\author{
  \\
  { Sergio Caracciolo}              \\
  {\small\it Dipartimento di Fisica and INFN, Universit\`a di Milano, 
    and INFM-NEST, }  \\[-0.2cm]
  {\small\it I-20133 Milano, ITALY}          \\[-0.2cm]
  {\small Internet: {\tt sergio.caracciolo@mi.infn.it}}   
  \\[-0.1cm]  \and
  { Andrea Gambassi }              \\
  {\small\it Max-Planck-Institut f\"ur Metallforschung,}  \\[-0.2cm]
  {\small\it Heisenbergstr. 3, D-70569, Stuttgart, GERMANY,}      \\[-0.2cm]
  {\small\it and Institut f\"ur Theoretische und Angewandte Physik,} \\[-0.2cm]
  {\small\it Universit\"at Stuttgart, Pfaffenwaldring 57, 
       D-70569 Stuttgart, GERMANY} \\[-0.2cm]
  {\small Internet: {\tt gambassi@mf.mpg.de}}   
  \\[-0.1cm]  \and
  { Massimiliano Gubinelli}              \\
  {\small\it Dipartimento di Matematica Applicata  and INFN,
       Universit\`a di Pisa} \\[-0.2cm]
  {\small\it I-56100 Pisa, ITALY}          \\[-0.2cm]
  {\small Internet: {\tt m.gubinelli@dma.unipi.it}}   
  \\[-0.1cm]  \and
  { Andrea Pelissetto}              \\
  {\small\it Dipartimento di Fisica and INFN,
           Universit\`a di Roma ``La Sapienza''}        \\[-0.2cm]
  {\small\it I-00185 Roma, ITALY}          \\[-0.2cm]
  {\small Internet: {\tt Andrea.Pelissetto@roma1.infn.it}}   \\[-0.2cm]
  {\protect\makebox[5in]{\quad}}  
  \\
}
\date{\today}

\begin{document}
\maketitle

\abstract{
We discuss the shape dependence of the finite-size scaling limit 
in a strongly ani\-so\-tropic $O(N)$ model in the large-$N$ limit. 
We show that scaling is observed even if an incorrect value for the 
 anisotropy exponent is 
considered.  However, the related exponents may only be effective ones, 
differing from the correct critical exponents of the model. 
We discuss the implications of our results for numerical finite-size 
scaling studies of strongly anisotropic systems.
}
\clearpage

\section{Introduction}

Finite-size scaling (FSS)~\cite{Fisher-71,Barber,Cardy-88,Privman-91} 
is a very powerful tool that allows to extract
information on the critical behavior of a system---which in
principle can only be observed in the infinite-volume limit---from
finite-volume results. In particular, the most recent Monte Carlo
studies heavily rely on FSS for the determination of critical
properties (see, e.g.,
Refs.~\cite{Binder-81,LWW-91,BLH-95,Caracciolo95,Caracciolo95b,%
HPV-99,BFMMPR-99,BST-99,CHPRV-02}
for recent applications to the $N$-vector model in two and three dimensions;
the list is of course far from being exhaustive). 
For isotropic systems the recipe is well
known. For instance, in two dimensions one considers a
domain $L\times M$ and the finite-size scaling 
limit $\xi_{L,M}\to\infty$, $L,M\to\infty$
at fixed $\xi_{L,M}/L$ and aspect ratio $L/M$. Here $\xi_{L,M}$ is
a suitably defined finite-volume correlation length. In this limit,
long-distance quantities, e.g. the susceptibility,
show a scaling behavior. For instance, if
$\cal O$ diverges in the thermodynamic limit
 as $|t|^{-x_{\cal O}}$ for $t\equiv T - T_c\to 0$
($T$ is the temperature and $T_c$ its critical value), then in the FSS
limit one finds
\begin{equation}
{\cal O}(t,L,M) \approx L^{x_{\cal O}/\nu} f_{\cal O} (\xi_{L,M}(t)/L,L/M),
\end{equation}
where $f_{\cal O}(x,y)$ is a universal function.

One may also ask what happens if one considers a different limit:
$\xi_{L,M}\to\infty$,
$L,M\to\infty$ keeping fixed the ratios $\xi_{L,M}/L$ and $L/M^{1+\delta}$
with $\delta\not= 0$. If $\delta < 0$, $M$ increases faster than
$L$ and it is easy to guess that we will obtain an effective
strip geometry so that
\begin{equation}
{\cal O}(t,L,M) \approx L^{x_{\cal O}/\nu} f_{\cal O} (\xi_{L,M}(t)/L,0).
\end{equation}
On the other hand, if $\delta > 0$, $M$ increases slowly, and, if we 
are able to keep
$\xi_{L,M}/L$ fixed, i.e.~to use $L$ as reference box size 
(it is not obvious that this is possible), the domain
effectively shrinks and becomes one-dimensional. Then, the question is 
whether
a scaling behavior is still observed. One may imagine that
the scaling function to be used is that of a one-dimensional system,
but in this case it is unclear which exponent should be used in the
prefactor.

This problem may appear academic at first but it is motivated by 
systems that are strongly anisotropic. For these
systems one introduces an anisotropy exponent $\Delta$ and the
{\em canonical} FSS limit is obtained by taking
$\xi_{L,M}\to\infty$, $L,M\to\infty$
keeping fixed $\xi_{L,M}/L$ and the aspect ratio $L/M^{1+\Delta}$.
Examples of such systems are provided by driven systems that admit
nonequilibrium stationary states with strong anisotropy~\cite{SchmZia,MD-99},
surface-growth processes~\cite{Krug}, 
Lifshitz points~\cite{Hornreich-80,Selke-92,PH-01,Diehl}, 
uniaxial magnets with dipolar
interactions~\cite{AF-73,BZ-76}, just to mention a few of them. 
A general approach to scale invariance in infinite volume for these 
anisotropic systems has been developed by Henkel ~\cite{Henkel}. 
However, FSS is still poorly understood. 
An exact computation on a dimer model which undergoes an 
anisotropic phase transition was performed by 
Bhattacharjee and Nagle~\cite{dimer}, while, in the context of the FSS for 
driven diffusive systems, a phenomenological FSS theory was proposed 
by Binder and Wang~\cite{binder} and by Leung~\cite{Leung-91,Leung-92}
who also gave some heuristic arguments on the consequences of taking the  
FSS limit with an incorrect aspect ratio.

Often in experimental or Monte Carlo applications the exponent $\Delta$ 
is not known {\em a priori}, and thus two questions naturally arise. 
First, if we
consider the FSS limit with fixed $L/M^{1+\Delta + \delta}$, $\delta \neq 0$,
do we still observe scaling? and if yes, with which exponents?
Second, how do we determine $\Delta$? An answer to these two questions
is of utmost practical importance.

In this work we analyze the finite-size scaling behavior of the
$N$-vector model in the  $N \to \infty$ limit, since in this case it
is exactly solvable, see, e.g., Ref.~\cite{ZinnBook}. 
This model has been a classical
example in FSS investigations, starting from the work of
Br\'ezin~\cite{Brezin82}, recently extended to several 
generalizations~\cite{Brankov},
including long-range interactions and boundary effects.
Here, we consider a general class of $O(\infty)$ models which includes
the classical isotropic short-range and long-range cases but also
models in which the spin-spin coupling decays with different power-laws
in different lattice directions, giving rise to a strongly anisotropic
phase transition. For these models we analyze the FSS limit
using finite boxes with arbitrary shape.

We show that $\Delta$ is uniquely determined if one properly measures the 
correlation length. On the other hand, if only zero-momentum quantities 
are available, for instance the susceptibility $\chi$, $\Delta$ cannot 
be easily determined. Indeed, even if the aspect ratio is incorrect, 
i.e. $\delta \not = 0$, one still observes scaling. In some cases, it is 
even possible to observe two different FSS limits with the {\em same}
data, one corresponding to the layer/strip geometry, the second one 
corresponding to a lower-dimensional system. 

Our analysis will be limited to systems below the upper critical 
dimension and above the lower critical one. Also, we will not address
the question of the FSS limit at fixed vanishing 
magnetization, which is of relevance for lattice-gas studies and has already
been discussed for isotropic systems in Ref.~\cite{CGGP-01}. 

The paper is organized as follows. In Sec.~\ref{sec2} we define the model,
the basic observables, and discuss the large-$N$ limit. 
In Sec.~\ref{sec3} we derive the anisotropy exponent $\Delta$ that 
defines the aspect ratio and report the FSS functions for the susceptibility
and the correlation lengths. In Sec.~\ref{sec4} we discuss the 
noncanonical FSS limit in which the ratio $M/L^{1+\Delta+\delta}$, 
$\delta \not= 0$, is kept 
fixed as the size of the lattice is increased. The results of these two 
sections are derived in the Appendices. In Sec.~\ref{sec5} we present a simple 
numerical example: we consider the standard isotropic model with short-range
interactions on a cubic lattice $M\times L^2$ and show that {\em two} 
different scaling behaviors can be observed by keeping fixed the 
ratio $M/L^{3/2}$, in agreement with the theoretical results of Sec.~\ref{sec4}.
Finally, in Sec.~\ref{sec6} we present our conclusions and discuss the 
implications for numerical studies. In particular, we discuss how 
one can determine numerically the anisotropy exponent $\Delta$.

\section{The model} \label{sec2}

We consider a $d$-dimensional hypercubic lattice $\mathbb{Z}^d$, 
unit $N$-vector spins $\bsigma$ defined at the sites of the lattice,
and the Hamiltonian
\begin{equation}
{\cal H} =\, - N \sum_{\pos{x},\pos{y}} J(\pos{x}-\pos{y}) 
\bsigma_\pos{x} \cdot \bsigma_\pos{y}  -
                N h \sum_{\pos{x}} \sigma_\pos{x}^1.
\label{eq2.1}
\end{equation}
The partition function is simply
\begin{equation}
Z =\, \int \prod_\pos{x} [d\bsigma_\pos{x} \, 
\delta(\bsigma_\pos{x}^2-1)] \; e^{-\beta{\cal H}},
\end{equation}
with $\beta \equiv 1/T$.
We will be interested in studying the finite-size behavior of the theory.
For this purpose, we consider a finite box $\Lambda_V$
of finite extent $M$ in the first $q$ directions (called the 
``parallel''
directions and denoted by the subscript $\|$) and $L$ in the remaining $p$
directions (called ``transverse'' and denoted by $\bot$), with $d=q+p$,
and therefore of volume $V=M^q L^p$. In order to be able to consider 
long-range interactions, we define a finite-size coupling
$J_{M,L}(\pos{x})$ as 
\begin{equation}
J_{M,L}(\pos{x}) = 
   \sum_{\pos{n}_\|\in \mathbb{Z}^p} \sum_{\pos{n}_\bot\in \mathbb{Z}^q}
    J(\pos{x} + \pos{n}_\| M + \pos{n}_\bot L),
\label{coupling-def}
\end{equation}
and the finite-size Hamiltonian
\begin{equation}
{\cal H} =\, - N \sum_{\pos{x},\pos{y}\in \Lambda_V} J_{M,L}(\pos{x}-\pos{y}) 
\bsigma_\pos{x} \cdot \bsigma_\pos{y}  -
                N h \sum_{\pos{x}\in \Lambda_V} \sigma_\pos{x}^1,
\label{eq2.1a}
\end{equation}
with periodic boundary conditions. Note that definition 
(\ref{coupling-def}) implies for the Fourier transforms
(with $\pos{p}\in\Lambda_V^*$, see Eq.~(\ref{duallatt}) below)
\begin{equation}
\widehat{J}_{M,L}(\pos{p}) = 
     \sum_{\pos{x}\in \Lambda_V} e^{i\pos{p}\cdot\pos{x}} J_{M,L}(\pos{x}) = 
     \sum_{\pos{x}\in \mathbb{Z}^d}  e^{i\pos{p}\cdot\pos{x}} J(\pos{x}) =
     \widehat{J}(\pos{p}).
\end{equation}
We consider anisotropic long-range interactions and thus we assume 
that asymptotically $\widehat{J}(\pos{q})$ has the form
\begin{equation}
\widehat{J}(\pos{q}) \simeq \widehat{J}(\pos{0})+  
a_\bot |\pos{q}_\bot|^{2\rho} +
a_\| |\pos{q}_\||^{2\sigma}, \qquad \text{for $|\pos{q}| \to 0$},
\label{Jq-smallq}
\end{equation}
with $0 < \rho,\sigma \le 1$ and 
$a_\bot < 0$, $a_\| < 0$ in order to have a ferromagnetic system. 
For simplicity, we will assume the two metric
factors to be equal, and, by redefining the inverse temperature $\beta$,
we can set $a_\bot = a_\| = -1/2$. 

In the large-$N$ limit,
assuming periodic boundary conditions and $h=0$,
the theory is solved in terms of the gap equations 
\begin{eqnarray}
\lambda_V \sigma_V = 0, \qquad\qquad
\beta = \beta \sigma_V^2 +
    {1\over L^p M^q} \sum_{\pos{q}\in \Lambda^*_V}
   {1\over K(\pos{q}) + \lambda_V },
\label{infty}
\end{eqnarray}
where $K(\pos{q}) = - 2 (\widehat{J}(\pos{q}) - \widehat{J}(\pos{0}))$ 
and $\Lambda^*_V$ is the
lattice
\begin{equation}
\Lambda_V^* = \left( {2\pi M^{-1}}\, \mathbb{Z}^q_{M}, {2\pi L^{-1}}\, 
\mathbb{Z}^p_{L}, \right).
\label{duallatt}
\end{equation}
In infinite volume, the same equations holds, with the simple
substitution of the summation with the normalized integral over the 
first Brillouin zone $[-\pi,\pi]^d$.

The meaning of the parameters $\lambda_V$ and $\sigma_V$ is clarified by
considering the magnetization and the two-point function.
If $\langle\cdot\rangle_V$ is the mean value for a system of volume 
$V$, we define
\begin{equation}
M_V =\, \langle \sigma^1\rangle_V, \qquad\qquad G_V(\pos{x}) = \langle 
\bsigma_\pos{0}\cdot\bsigma_\pos{x}\rangle_V.
\end{equation}
Then
\begin{equation}
M_V = \sigma_V, \qquad\qquad \widehat{G}_V(\pos{q}) = {\beta^{-1}\over 
K(\pos{q}) + \lambda_V},
\label{prop}
\end{equation}
where $\widehat{G}(\pos{q})$ is the Fourier transform of $G(\pos{x})$.
As we show in App.~\ref{AppB}
in the limit $V\to \infty$ and in the scaling limit $\lambda_V\to 0$,
the correlation function has the  form
\begin{equation}
G_\infty(\pos{x}) = \xi_\bot^{\rho(2-D)} \widetilde G_\infty 
(\pos{x}_\bot/\xi_{\bot,\infty}, \pos{x}_\|/\xi_{\|,\infty}),
\end{equation}
where
\begin{equation}
   \xi_{\bot,\infty} = \lambda_\infty^{-1/2\rho}, \qquad \xi_{\|,\infty} 
= \lambda_\infty^{-1/2\sigma}.
\end{equation}
Therefore, even if the model has long-range correlations in the high-$T$ phase,
it is sensible to look at $\xi_{\bot,\infty}$ and $\xi_{\|,\infty}$ as
appropriate typical length scales of the system. We 
will refer to them respectively as transverse and longitudinal 
correlation lengths.
They are related by 
\begin{equation}
\xi_{\|,\infty} = \xi_{\bot,\infty}^{\rho/\sigma}.
\label{xi-relation}
\end{equation}
The critical point is characterized by a vanishing mass gap,
i.e. $\xi_{\bot,\infty}^{-1}= \xi_{\|,\infty}^{-1} = 0$, 
and by a vanishing magnetization, $\sigma_\infty=0$, so that the
critical temperature is given by (see Eq.~\reff{infty})
\begin{equation}
\beta_c = \int_{[-\pi,\pi]^{d}} \frac{d^d
   \pos{p}}{(2\pi)^d} \frac{1}{K(\pos{p})},
\label{betacrit}
\end{equation}
which is finite whenever the effective dimensionality $D \equiv
p/\rho+q/\sigma$ is greater than $2$ and infinite for $D\le 2$ (the 
system undergoes a
zero-temperature phase transition).
Moreover, given that we will be interested in finite-volume properties,
we have $M_V=0$ for all values of $\beta$, so that we can set 
$\sigma_V=0$ in the gap equation.

\section{Canonical finite-size scaling} \label{sec3}

We do not address here the problem of the definition of a
finite-volume correlation length and we simply use 
\begin{equation}
   \xi_{\bot,V} = \lambda_V^{-1/2\rho}, \qquad \xi_{\|,V} = 
\lambda_V^{-1/2\sigma},
\end{equation}
which are the finite-volume analogues of the infinite-volume correlation 
lengths defined above.

In order to perform the FSS limit we must identify the correct aspect ratio.
{}From Eq.~(\ref{Jq-smallq}) we see that $q_\bot^\rho\sim q_\|^\sigma$, i.e.
$L^{-\rho} \sim M^{-\sigma}$. Therefore, the FSS limit should be taken 
keeping $S\equiv M/L^{1+\Delta}$, $\Delta \equiv \rho/\sigma - 1$, fixed. 
In App.~\ref{AppA} we show that, in the limit 
$L,M,\xi \to \infty$ with $S$ and $\xi_{\bot,V}/L$ (or, equivalently,
$\xi_{\|,V}/M$)
fixed, below the upper critical dimension,
i.e. for $D < 4$, the gap equation takes a scaling form. 
Above the lower critical dimension, i.e. for $D > 2$, we obtain
\begin{equation}
\label{eq:gap_equation_sconst}
(4\pi)^{\rho(2-D)/2} (\beta-\beta_c)L^{\rho(D-2)} z^{2-D} =
- A_{p,q,\rho,\sigma} + I_{p,q,\rho,\sigma}(z,S),
\end{equation}
where $z\equiv (4\pi)^{-\rho/2} (L/\xi_{\bot,V})^\rho$,
$\beta_c$ is given in Eq.~\reff{betacrit}, 
\begin{equation}
A_{p,q,\rho,\sigma} \equiv - \frac{1}{(4\pi)^{(p+q)/2}}\frac{1}{\rho\,\sigma}
\frac{\Gamma(p/2\rho)\Gamma(q/2\sigma)}{\Gamma(p/2)\Gamma(q/2)} 
   \Gamma\left(1-{p\over 2\rho} - {q\over 2\sigma}\right) ,
\label{def-Apq}
\end{equation}
and $I_{p,q,\rho,\sigma}(z,S)$ is defined in App.~\ref{AppA},
cf. Eq.~(\ref{sigma-as}). 

For $D<2$ there is no finite-temperature phase transition and the gap equation
assumes the form
\begin{equation}
\label{eq:gap_equation_sconst2}
(4\pi)^{\rho(2-D)/2}  \beta L^{\rho(D-2)} z^{2-D} =
- A_{p,q,\rho,\sigma} + I_{p,q,\rho,\sigma}(z,S).
\end{equation} 
Considering the case $D>2$, 
Eq.~(\ref{eq:gap_equation_sconst}) shows that
\begin{equation}
{\xi_{\bot,V}\over L} = f_{\xi,1}[
   (\beta - \beta_c) L^{\rho(D-2)},S] = 
   f_{\xi,2}[
   (\beta - \beta_c) M^{\sigma(D-2)},S] ,
\end{equation}
that allows us to define a ``parallel" exponent $\nu_\| $ and a 
``perpendicular" exponent $\nu_\bot$ as 
\begin{equation}
 \nu_\bot = {1\over \rho(D-2)}, \qquad
 \nu_\| = {1\over \sigma(D-2)}, \qquad
 {\nu_\| \over \nu_\bot} = 1 + \Delta = {\rho\over \sigma}.
\end{equation}
Analogously, for the finite-volume susceptibility 
$\chi_{V} = \beta \widehat G_{V}(0)$ we obtain 
\begin{equation}
\chi_{V} = \lambda_V^{-1} = L^{2\rho} f_{\chi,1}[
   (\beta - \beta_c) L^{\rho(D-2)},S] =
   M^{2\sigma} f_{\chi,2}[
   (\beta - \beta_c) M^{\sigma(D-2)},S].
\label{chi-scaling}
\end{equation}
The corresponding critical exponent $\gamma$ associated with the behavior in
infinite volume, $\chi_\infty\sim (\beta_c - \beta)^{-\gamma}$,
is correctly identified as 
$\gamma = 2 \rho \nu_\bot = 2 \sigma \nu_{\|} = 2(D-2)^{-1}$.

\section{Noncanonical finite-size scaling} \label{sec4}

\subsection{General considerations} \label{sec4.1}

In Sec. \ref{sec3} we discussed the canonical FSS, obtained by
keeping $S$ constant. However, in many anisotropic systems the correct
exponent $\Delta$ is not known and thus an important question is
what happens if we consider the FSS limit keeping fixed the ratio
\begin{equation}
S_\delta \equiv \frac{M}{L^{1+\Delta+\delta}}
\label{anomalous-scaling}
\end{equation}
with $\delta\neq 0$. If one has separately defined parallel and 
transverse correlation lengths, one can immediately identify the 
correct anisotropy exponent $\Delta$. It corresponds to the value for 
which $\xi_{\bot,V}/L$ and $\xi_{\|,V}/M$ both remain finite in the 
FSS limit. This uniquely defines the correct exponent. This is not 
surprising, since the correlation lengths satisfy Eq.~(\ref{xi-relation})
that essentially defines the correct aspect ratio. On the other hand,
one may not have access to the correlation lengths but only to some 
zero-momentum correlation function, for instance to the susceptibility. 
Thus, one may ask what kind of scaling behavior, if any, is observed 
if $S_\delta$ is kept fixed. In other words, we may ask if 
we can find exponents $\tilde{\gamma}$, 
$\tilde{\nu}_\bot$, and $\tilde{\nu}_\|$ so that asymptotically,
for $L$ and $M$ going to infinity and $\beta\to\beta_c$, we observe a scaling
behavior analogous to that defined in Eq.~(\ref{chi-scaling}), i.e.
\begin{eqnarray}
\chi &=& L^{\tilde{\gamma}/\tilde{\nu}_\bot} 
   f_{\chi,1}[(\beta - \beta_c) L^{1/\tilde{\nu}_\bot},S_\delta] 
\label{chieff-L} \\
     &=& M^{\tilde{\gamma}/\tilde{\nu}_\|} 
   f_{\chi,2}[(\beta - \beta_c) M^{1/\tilde{\nu}_\|},S_\delta].
\label{chieff-M} 
\end{eqnarray}
Of course, 
the second important question is the relation of these exponents
with the correct ones defined above. Again, we will restrict our attention
to the case $2 < D < 4$, i.e.~ below the upper critical dimension and 
above the lower critical one.

In the following, we only consider the case $\delta>0$, given that
the case $-1-\Delta<\delta<0$ is obtained by performing
the substitutions 
\begin{equation}
  p\leftrightarrow q,\quad L\leftrightarrow M,\quad
\rho \leftrightarrow \sigma, \quad
\delta \mapsto -{\delta \sigma^2/\rho(\rho + \sigma \delta)}, \quad
S_\delta \mapsto S_\delta^{-1/(\rho/\sigma+\delta)}.
\end{equation}
If $\delta$ is positive, $S$ goes to infinity as $L, M \to\infty$. 
We must then study the gap 
equation in that limit, requiring at the same time $\lambda_V\to 0$ 
in an arbitrary way. The detailed calculation is presented in 
App.~\ref{AppC}, where we show that there are two nontrivial cases:
\begin{itemize}
\item[(a)]
One can take the limit $S\to \infty$ at $z^2\sim\lambda_V L^{2\rho}$ fixed. 
In terms of the correlation lengths this corresponds to the FSS limit 
at $\xi_{\bot,V}/L$ fixed. At the same time 
$\xi_{\|,V}/M \sim (\xi_{\bot,V}/L)^{\rho/\sigma} L^{-\delta}$ goes to 
zero, i.e. $M$ increases faster than parallel correlations. Therefore,
the final ``effective'' geometry is $\infty^q \times L^p$, so that 
we obtain the FSS behavior of a ``layer.''
\item[(b)]
Alternatively, one can take the limit $S\to \infty$, with $z\to 0$,
keeping $z^2 S^{2\sigma}\sim \lambda_V M^{2\sigma}$ fixed. This corresponds to 
keeping fixed the ratio $\xi_{\|,V}/M$ while at the same time 
$\xi_{\bot,V}/L$ diverges. The expected ``effective''
geometry should be that of a $q$-dimensional hypercube with linear
size  $M$ ($M^q \times 0^p$).
\end{itemize}
The first limit always exists and it is easy to predict the result.
It correspond to the standard FSS limit for a system $\infty^q \times L^p$
and thus the scaling in terms of $L$ is canonical. Therefore,
it corresponds to taking $\beta\to \beta_c$ and 
$L \to \infty$ keeping fixed $(\beta - \beta_c) L^{\rho (D-2)}$, which 
is indeed the correct combination. Then, we obtain 
\begin{equation}
\chi = L^{2\rho} f_{\chi,1}[
   (\beta - \beta_c) L^{\rho(D-2)},\infty]
\label{chi-scal-(a)-L}
\end{equation}
and, using Eq.~(\ref{anomalous-scaling}), 
\begin{equation}
\chi = \left(M\over S_\delta\right)^{{2\rho\over 1 + \Delta + \delta}}
f_{\chi,1}\left[
   (\beta - \beta_c) 
   \left(M\over S_\delta\right)^{{\rho (D-2)\over 1 + \Delta + \delta}},
         \infty\right]\; .
\label{chi-scal-(a)-M}
\end{equation}
Thus $\tilde{\nu}_\bot = \nu_\bot$ and $\tilde{\gamma} = \gamma$ (note that 
both equations give the same result for $\gamma$).
On the other hand,
the parallel exponent ${\tilde\nu}_\|$ is given by
\begin{equation}
{\tilde\nu}_\| = {1 + \Delta + \delta\over 1 + \Delta}  \nu_\|\; ,
\end{equation}
a result which is a trivial consequence of the fact that if $S_\delta$
is fixed, then ${\tilde\nu}_\|/{\tilde\nu}_\bot = 1 + \Delta + \delta$.
Therefore, even if $\delta \not = 0$, one still observes scaling. 
Transverse exponents are correct, while longitudinal ones are only effective.
Let us finally note that Eqs.~(\ref{chi-scal-(a)-L}) and 
(\ref{chi-scal-(a)-M}) give both the correct infinite-volume behavior 
for $\chi$. Indeed, the existence of a finite infinite-volume limit 
for $\beta < \beta_c$ implies $f_{\chi,1}(x) \sim (-x)^{-2/(D-2)} \sim 
(-x)^{-\gamma}$ for $x\to-\infty$ and therefore 
 Eqs.~(\ref{chi-scal-(a)-L}) and (\ref{chi-scal-(a)-M}) 
are consistent with $\chi_\infty(\beta) \sim (\beta_c - \beta)^{-\gamma}$.
In other words, one must have $\tilde{\gamma} = \gamma$ in order to 
have the correct infinite-volume limit.

Limit (b) is much less conventional and 
we will show that such a limit exists (in the sense that 
$\chi$ has a scaling behavior) only if $q/\sigma < 2$, i.e.
if the $q$-dimensional theory is below the lower critical dimension:
in this case however, the exponents obtained from the data collapse do not have 
anything to do with the correct ones, while the scaling functions
depend only on the $q$-dimensional theory, i.e. theories with different 
$p$ and/or $\rho$---they are thus physically inequivalent---have the 
{\em same} scaling functions. Thus, in this case from the observation of 
a good scaling behavior one can draw the incorrect conclusion that 
physically inequivalent theories have the same critical behavior.

At this point the reader may be puzzled by the fact that for the same 
$S_\delta$ we observe two different types of scaling. 
Mathematically this is related to the fact that in the FSS limit we must follow 
a path in the $\beta$, $L$ plane with $\beta\to\beta_c$ and $L\to\infty$. 
The two limits correspond to two different families of paths. 
Unexpectedly scaling is observed on both of them.

\subsection{Limit (b)} \label{sec4.2}

We wish now to consider the limit $S\to \infty$, $z\to 0$ at fixed 
$z S^\sigma$. Such a limit has been considered in App.~\ref{AppC} where it 
is shown that the leading contribution to the gap equation has the form
\begin{eqnarray}
(4\pi)^{\rho(2-D)/2+d/2} (\beta-\beta_c) L^{\rho (D-2)} = 
   (4\pi)^{q/2} r^{\sigma - q/2} \omega^{-2+q/\sigma} 
   [I_{q,\sigma}^{\rm iso}(\omega) - A_{q,\sigma}] + K,
\label{gapeq-limitb}
\end{eqnarray}
where 
$r \equiv (4\pi)^{\rho/\sigma-1} S^2$, 
$\omega \equiv z r^{\sigma/2}  = (4 \pi)^{-\sigma/2} \lambda_V^{1/2} M^\sigma$, 
$A_{q,\sigma} \equiv A_{0,q,\rho,\sigma}$ (it is easy to see that it 
is $\rho$-independent), 
and $I_{q,\sigma}^{\rm iso}(\omega)$ is the analogue of 
$I_{p,q,\rho,\sigma}(z,S)$ appearing in 
Eqs.~(\ref{eq:gap_equation_sconst}) and (\ref{eq:gap_equation_sconst2}) 
for an isotropic $q$-dimensional system with long-range exponent $\sigma$.
The constant $K$, cf.~Eq.~(\ref{def-K}),
gives a subleading correction for $q/\sigma< 2$. 
We will now distinguish two cases, depending on the value of 
$q/\sigma$. 

\subsubsection{Case $0<q/\sigma<2$} \label{sec4.2.1}

In this case we can neglect $K$ and set $r = r_\delta 
L^{2\delta}$, where 
$r_\delta \equiv (4\pi)^{\rho/\sigma-1} S^2_\delta$  (by definition $r_0=r$) 
is kept fixed in the limit. Then, for $\beta > \beta_{c}$, we can rewrite 
\begin{eqnarray}
(4\pi)^{\rho(2-D)/2+p/2} r_\delta^{q/2-\sigma}
   (\beta-\beta_c) L^{\rho (D-2) - 2\delta(\sigma-q/2)}  \omega^{2-q/\sigma} = 
   I_{q,\sigma}^{\rm iso}(\omega) - A_{q,\sigma}.
\label{scaling-anomalo}
\end{eqnarray}
This equation has the same functional form as the gap equation of an
isotropic $q$-dimensional system with effective dimension 
$d_{\rm eff} = q/\sigma < 2$, 
cf. Eq.~(\ref{eq:gap_equation_sconst2}). Therefore, if for a $q$-dimensional 
isotropic system of extent $M$ and long-range exponent $\sigma$ we have,
\begin{eqnarray}
\chi^{\rm iso} = M^{2\sigma} f^{\rm iso}_{\chi} 
    [M^{\sigma (d_{\rm eff} - 2)} \beta],
\end{eqnarray}
then in the limit we are considering we obtain
\begin{eqnarray}
\chi = M^{2\sigma} f^{\rm iso}_{\chi} 
[S_{\delta}^{-p/(1+\Delta+\delta)} M^{1/{\tilde\nu_\|}} (\beta - \beta_c)] ,
\label{chi-newscaling}
\end{eqnarray}
where 
\begin{equation}
{\tilde\nu}^{-1}_\| = 
   {(1+\Delta)\nu^{-1}_\| + \delta\sigma(d_{\rm eff} - 2)\over 1+\Delta+\delta}.
\label{nupareffective}
\end{equation}
We note that the second term in the numerator of
Eq.~(\ref{nupareffective}) is equal to $\delta\, \nu^{-1}_{\|,{\rm iso}}$,
where $\nu_{\|,{\rm iso}} = [\sigma(d_{\rm eff} - 2)]^{-1}$ 
is the correct  exponent for the
$d_{\rm eff}$-dimensional isotropic system, so that 
Eq.~(\ref{nupareffective}) interpolates between 
${\nu}^{-1}_\|$ and $\nu^{-1}_{\|,{\rm iso}}$.

There are two limitations to the validity of Eq.~(\ref{chi-newscaling}).
First,
Eq.~(\ref{chi-newscaling}) holds only in the limit in which the 
argument is constant as $M\to \infty$. Since we are considering 
$\beta\to \beta_c$, this requires ${\tilde\nu}_\| > 0$ and in turn
\begin{equation}
\delta < \delta_{\rm max} = {(D-2) (1 + \Delta)\over (2 - d_{\rm eff})} .
\end{equation}
If $\delta$ is larger than $\delta_{\rm max}$ no scaling is observed 
by taking $\beta\to\beta_c$. One observes a truly $q$-dimensional 
FSS behavior: a scaling behavior is obtained only by taking $\beta\to\infty$.

Additionally, it should be observed that $f_\chi^{\rm iso}(x)$ is defined 
only for $x>0$, and indeed a solution to Eq.~(\ref{scaling-anomalo})
is found only when the left-hand side is positive. 
In Appendix \ref{AppC} we show that the correct extension for $x < 0$  
is $f_\chi^{\rm iso}(x) = 0$ in the sense that, for $\beta < \beta_c$, we have 
$\chi/M^{2\sigma}\to 0$.

We also obtain 
\begin{eqnarray}
\tilde{\gamma} &=& 2 \sigma \tilde{\nu}_\|, \nonumber \\
\tilde{\nu}_\bot &=& {\tilde{\nu}_\|\over 1+\Delta+\delta}.
\end{eqnarray}
The effective exponents are thus unrelated to the correct ones
and vary continuously with $\delta$, interpolating between the 
correct ones ($\delta = 0$) and those of a $d_{\rm eff}$-dimensional
system ($\delta = \delta_{\rm max}$).

At this point the reader may be puzzled by the fact that 
$\tilde\gamma\not=\gamma$, since we already claimed in Sec.~\ref{sec4.1}
that this is a necessary condition to ensure the correct infinite-volume
limit. We now show that there is no contradiction. The effective 
exponent $\tilde\gamma$ differs from $\gamma$ because in limit (b) 
the scaling function $f_\chi(x)$ vanishes identically in the whole 
high-temperature phase $x<0$, i.e. in the only phase in
which $\chi$ is finite ($\chi_\infty(\beta) = \infty$ in the whole 
low-temperature phase). In other words, scaling (b) does not give any
information on the infinite-volume behavior, and thus there is no
surprise for $\tilde\gamma\not=\gamma$.

\subsubsection{Case $q/\sigma>2$} \label{sec4.2.2}

In this case $K$ in Eq.~(\ref{gapeq-limitb}) cannot be neglected,
and therefore no scaling solution can be found. 
Scaling can only be observed along a very specific trajectory 
in the $\beta,L$ plane. Indeed, if we consider the line
\begin{equation}
    \beta_c(L) = \beta_c + a L^{-1/\nu_\bot} ,
\label{betaL-def}
\end{equation}
where 
\begin{equation} 
    a = (4\pi)^{-\rho(2-D)/2-d/2} K ,
\end{equation}
we still observe scaling in the sense that asymptotically
\begin{equation}
\chi = S_\delta^{2\sigma} L^{2\sigma(1+\Delta+\delta)} f_\chi^{{\rm iso}}
   [S_\delta^{-2\sigma +q} L^{1/\tilde{\nu}_\bot}(\beta-\beta_c(L))],
\end{equation}
where $\tilde{\nu}_\bot^{-1} = \nu_\bot^{-1} + 
\delta \sigma (d_{\rm eff} -2) = \nu_\bot^{-1} + \delta\nu_{\|,\rm iso}^{-1}$ 
and $f_\chi^{{\rm iso}}$ is the scaling 
function of a $d_{\rm eff}$-dimensional isotropic system (with
$d_{\rm eff}=q/\sigma>2$).
However, this is not standard scaling since $a$ must be tuned to a proper
value: it is not enough to take $\beta\to\beta_c$, but it is also
necessary to take this limit along the line (\ref{betaL-def}).

\section{A numerical example} \label{sec5}

We wish now to present a numerical example, in order to clarify the 
issues discussed in the previous sections. We consider the standard 
short-range model with nearest-neighbor couplings, so that 
$\rho = \sigma = 1$ and $\Delta = 0$. We consider a three-dimensional 
cubic lattice of dimensions $M\times L^2$, i.e. $p=2$ and $q=1$. For this
system \cite{BZ-92}
\begin{equation}
\beta_c = {\sqrt{3} - 1\over 192 \pi^3} 
      \Gamma(1/24)^2 \Gamma(11/24)^2 \approx 
      0.25273100985866300303\; .
\end{equation}
According to the results of Sec.~\ref{sec4.2} we should be able to observe 
a noncanonical scaling for $0 < \delta < 1$. We thus fix $\delta = 1/2$
and compute $\chi$ for several values of $\beta$, $L$, and $M$ 
corresponding to $S_{1/2} = 1$ and $S_{1/2} = 10$. We use 
$5 \le L \le 245$ and $0.235 \le \beta \le 0.279$. 
The previous results predict two possible scalings. Scaling (a)
corresponds to consider $\chi/M^{4/3}$ vs $(\beta - \beta_c) M^{2/3}$, 
while scaling (b) corresponds to consider $\chi/M^{2}$ 
vs. $(\beta - \beta_c) M^{1/3}$. 

\begin{figure}
\epsfig{file=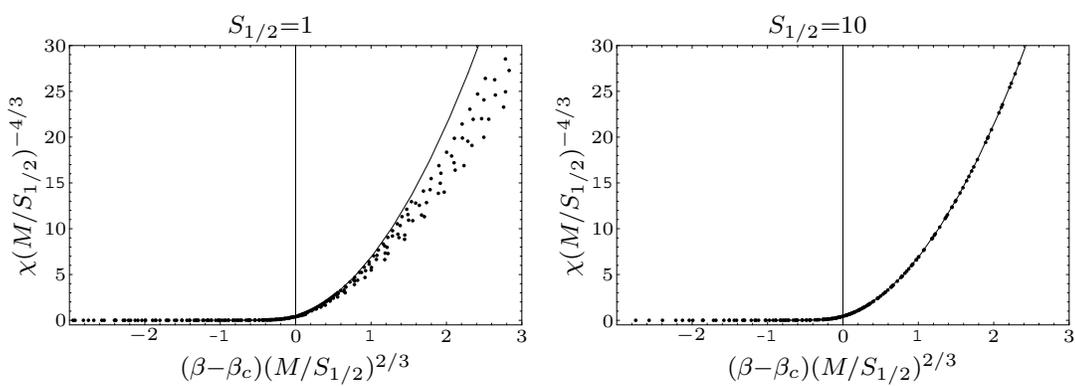,width=\textwidth}
\caption{Noncanonical FSS scaling in 3 dimensions: limit (a). The solid line 
is the theoretical prediction. Here $\delta = 1/2$. }
\label{fig:anomalous1}
\end{figure}

\begin{figure}
\epsfig{file=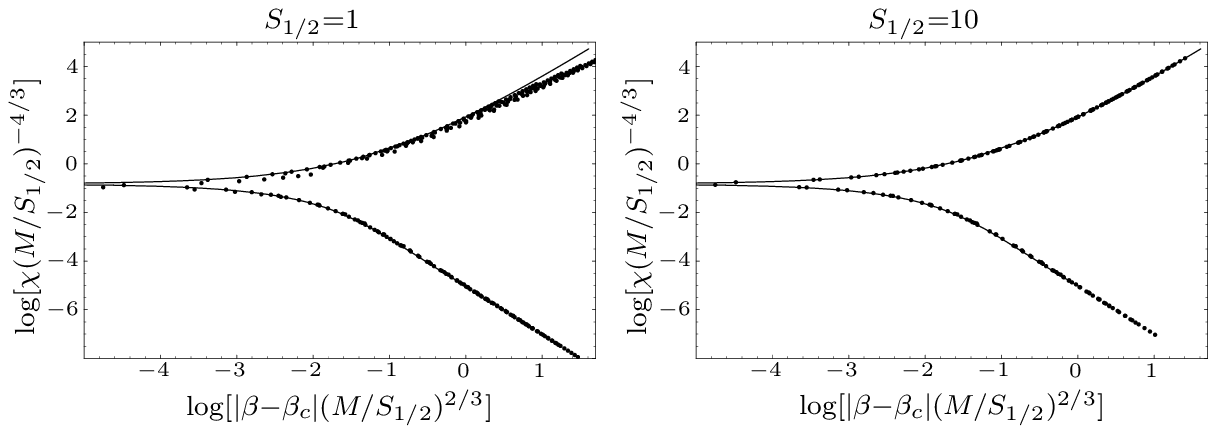,width=\textwidth}
\caption{Noncanonical FSS scaling in 3 dimensions: limit (a). The same data 
of Fig.~\protect\ref{fig:anomalous1} in a log-log plot.}
\label{fig:anomalous1-log}
\end{figure}

In Fig.~\ref{fig:anomalous1} we report the results for limit (a). 
For $S_{1/2} = 1$ there are somewhat large corrections to scaling 
and the convergence is slow in the low-temperature region. 
For $S_{1/2} = 10$ the agreement is very good: no corrections can be seen 
on this scale. This different behavior is due to the different lattice sizes 
that are used: the data have the same values of $L$ and therefore 
lattices with $S_{1/2} = 10$ are ten times larger than those with 
$S_{1/2} = 1$. In Fig.~\ref{fig:anomalous1-log} we show the same data 
in a log-log plot. Here we see better the results corresponding to
the high-temperature phase, although corrections in the low-temperature
phase are somewhat less visible.

\begin{figure}
\epsfig{file=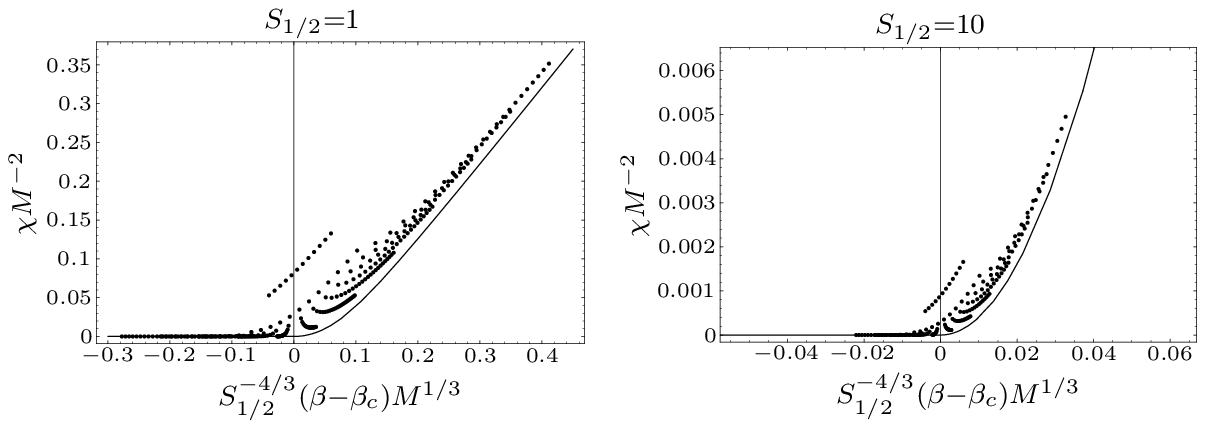,width=\textwidth}
\caption{Noncanonical FSS scaling in 3 dimensions: limit (b). The solid line
is the theoretical prediction. Here $\delta = 1/2$. }
\label{fig:anomalous2a}
\end{figure}

\begin{figure}
\epsfig{file=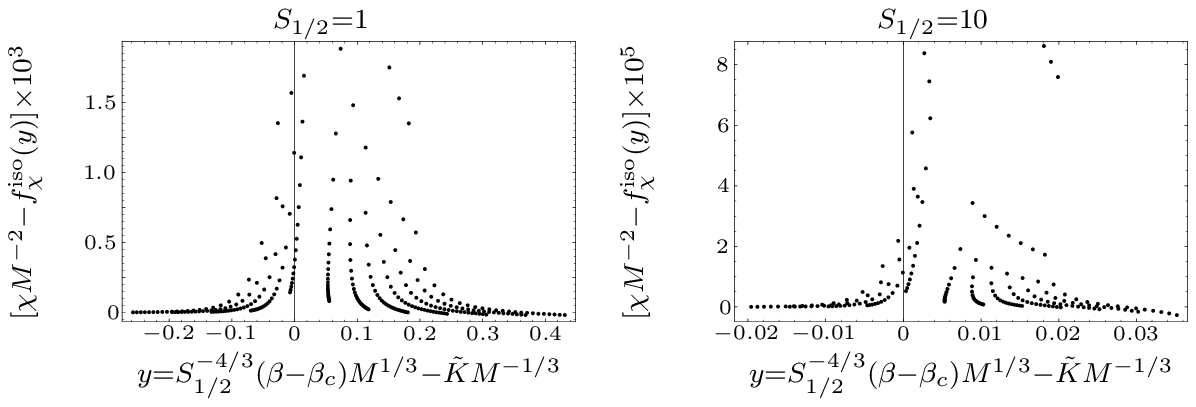,width=\textwidth}
\caption{Corrections to the FSS behavior: limit (b). Here $\delta = 1/2$.}
\label{fig:anomalous2b}
\end{figure}

In Fig.~\ref{fig:anomalous2a} we report the results for limit (b). 
The data show a reasonably good scaling behavior, although the convergence to 
the limiting curve, corresponding to a one-dimensional system, is quite slow. 
For limit (b) we have also investigated the behavior of the leading correction 
term.  In App.~\ref{AppC.3} we compute the first scaling correction to 
Eq.~(\ref{chi-newscaling}). 
It can be effectively taken into account by a size-dependent 
shift of the critical temperature, i.e. by writing 
\begin{equation}
\label{eq:scaling-corr}
\chi M^{-2} = f_{\chi}^{\text{iso}}[S_{\delta}^{-4/3} (\beta-\beta_{c}) M^{1/3}
           - \tilde K M^{-1/3}] ,
\end{equation}
where $\tilde K \equiv K (4 \pi)^{-1} S_{\delta}^{-2/3}$, $K$ is 
defined in Eq.~(\ref{def-K}). Numerically, in this particular case, 
$K \approx -3.9002649200019558828$. 
In Fig.~\ref{fig:anomalous2b} we plot the difference between the left-hand and 
the right-hand side of Eq.~(\ref{eq:scaling-corr}) versus the argument of the 
function $f_{\chi}^{\text{iso}}$. It is clear that data points in 
Fig.~\ref{fig:anomalous2b} are converging towards zero, confirming that 
asymptotically $f_{\chi}^{\text{iso}}(x)$ is indeed the correct scaling
function. A numerical analysis shows that corrections to 
Eq.~(\ref{eq:scaling-corr}) vanish as $M^{-1}$.

From a practical point of view it is important to be able to distinguish 
scaling (a) from scaling (b). This is not very difficult since in scaling 
(b) $f_\chi(x)=0$ in the region $x<0$. This observation provides a simple 
criterion: if the high-temperature data scale onto a nontrivial curve 
we are clearly observing limit (a): for instance the scaling curve 
appearing in Fig.~\ref{fig:anomalous1-log} can only correspond to scaling
(a).

\section{Conclusions} \label{sec6}

In this paper we have considered an anisotropic $O(\infty)$ model,
focusing on the shape dependence of the FSS limit and in particular
on the role of the aspect ratio. Considering for instance the susceptibility,
we have shown that, even if the FSS limit is taken by keeping constant
an incorrect aspect ratio $M/L^{1+\Delta+\delta}$, one still observes scaling, 
i.e.~the
data satisfy the scaling forms~(\ref{chieff-L}) and (\ref{chieff-M}), 
although the effective exponents differ from the correct ones. Therefore, the 
numerical observation of data collapse does not give direct 
information on the correct values of the anisotropy exponent $\Delta$ 
and of the critical exponents. 
In the case of limit (a), if we consider the scaling at fixed 
$M/L^\alpha$, we obtain for $\alpha \ge 1 + \Delta$
\begin{eqnarray}
\tilde{\nu}_\bot &=& \nu_\bot  \nonumber \\
\tilde{\nu}_\|   &=& \alpha\nu_\bot   \nonumber \\
\tilde{\gamma}   &=&  \gamma \; ,
\label{limita-1}
\end{eqnarray}
while for $\alpha \le 1 + \Delta$
\begin{eqnarray}
\tilde{\nu}_\bot &=& \nu_\|/\alpha  \nonumber \\
\tilde{\nu}_\|   &=& \nu_\|   \nonumber \\
\tilde{\gamma}   &=&  \gamma \; .
\label{limita-2}
\end{eqnarray}
These expressions have been explicitly verified in the specific model we 
consider, 
but we expect them to be valid for generic anisotropic systems.
In the case of limit (b) we are not much interested in the specific form
of the effective exponents. The only important feature that we must notice 
is that all exponents vary continuously with $\alpha$ and, if $\alpha$ is 
sufficiently different from $1 + \Delta$, scaling is observed only for 
$\beta\to \infty$.
                     
From a practical point of view, it is very important to 
establish a strategy that can be used to correctly identify the exponents
in anisotropic systems.  One possibility consists 
in performing a numerical study keeping 
fixed $M/L^\alpha$ for different values of $\alpha$. If one 
observes that the data obtained with two different values of $\alpha$
collapse onto a single curve by using the {\em same} exponents and the 
{\em same} value of $\beta_c$, then the exponents 
which are determined in this way are correct. Indeed, this implies 
that one is observing limit (a)---in limit (b) all exponents vary 
continuously---and therefore Eqs.~(\ref{limita-1}) (\ref{limita-2}) apply.
Note that as an additional check one can see whether $f_\chi(x)$ is 
nontrivial for $x < 0$: if this the case, one is observing limit (a).
To be more specific, suppose that the 
scaling behavior (\ref{chieff-L}) is observed for $\alpha_1 $ and $\alpha_2$, 
$\tilde{\gamma}$ and $\tilde{\nu}_\bot$ being the same in both cases;
then $\tilde{\gamma}$ and $\tilde{\nu}_\bot$ can be identified with 
$\gamma$ and $\nu_\bot$ and $1 + \Delta \le {\rm min}\, (\alpha_1,\alpha_2)$. 
From the data at $\alpha_1 $ and $\alpha_2$ 
one cannot determine $\nu_\|$ and $\Delta$. For this 
purpose, one must perform simulations at a value $\alpha_3$ such that 
$\tilde{\nu}_\bot\not= \nu_\bot$, i.e.~one must find a value such 
that $\alpha_3 < 1 + \Delta$. Then, one should use Eq.~(\ref{chieff-M}) 
setting $\tilde{\gamma} = \gamma$ (this is meant to avoid limit (b))
and using $\tilde{\nu}_\|$ as a free parameter. Then $\nu_\| = \tilde{\nu}_\|$ 
and $1 + \Delta = \nu_\|/\nu_\bot$. As a check, one can analyze 
the data at $\alpha_1$ and $\alpha_2$ and look for the scaling
behavior~(\ref{chieff-M}), fixing $\gamma$ to avoid limit (b). 
The effective exponents $\tilde{\nu}_\|$ should satisfy 
$\tilde{\nu}_\| = \alpha {\nu}_\| /(1 + \Delta)$.
Thus, in principle, by studying the behavior for three different values 
of $\alpha$ one can determine all exponents. However, in practice 
this may be difficult, because it requires the ability to distinguish 
corrections which vanish as $M,L\to \infty$ from corrections
that persist in the limit $M,L\to \infty$ and that require therefore 
the use of different values of the effective exponents. 

From a practical point of view it is important to note that all the 
above-discussed ambiguities are not present if the correlation lengths
are measured, since their scaling behavior is fixed, 
$\xi_\bot \sim L$, $\xi_\| \sim M$. Deviations from these laws give immediately
the exponent $\Delta$. Therefore, finite-size 
numerical studies of anisotropic systems 
should always determine finite-volume correlation lengths, as done 
for instance in Ref.~\cite{CGGP-02} for a driven lattice gas. 

A somewhat unexpected result of our analysis is that in some cases it is 
possible to observe two different types of effective scaling:
one corresponding to 
the layer/strip geometry $\infty^q\times L^p$, which is the scaling 
one would naturally expect, and the other one corresponding to 
a lower-dimensional system $M^q \times 0^p$. The possibility of this type 
of scaling is not obvious since it requires that one correlation length 
increases much faster than the shortest of the two lattice sizes. 
Even less clear is why this limit exists only if $q$ is below the 
lower critical dimension of the model (for larger values of $q$ 
such a scaling can still be observed by performing an additional tuning).
We expect that scaling (b) is strictly related to the fact that 
$\chi$ and $\xi$ as defined 
here  diverge in the whole low-temperature region, but it is 
difficult to transform this conjecture into a quantitative argument.
In Ising systems it is simple to avoid this feature. 
For instance, one could define a finite-volume susceptibility as 
\begin{equation}
\chi = \beta (\widehat{G}(0) - V M^2),
\end{equation}
where 
\begin{equation} 
M  = {1\over V} \left \langle \left |\sum_{i\in\Lambda_V}
       \sigma_i\right|\right\rangle.
\end{equation}
An open question is whether such a quantity may show a noncanonical scaling (b).
Our guess is that only scaling (a) is possible in this case.

\appendix
\section{Gap equation: Finite-size scaling limit} \label{AppA}

In this Appendix we wish to determine the asymptotic form of the
gap equation for $\lambda_V\to 0$, $M,L\to \infty$, without
making any hypothesis on the way in which the limit is taken.
We will assume to be below the
upper critical dimension, i.e. $D<4$ where $D\equiv p/\rho + q/\sigma$.
We will consider separately the cases $D<2$ and $D>2$.

\subsection{Gap equation above the lower critical dimension: 
$2 < D < 4$} \label{AppA.1}

We start by rewriting (see Ref.~\cite{Brezin82}) the gap equation as
\begin{equation}
\label{eq:app-gapeq}
\beta - \beta_c = - \lambda_V \int_{[-\pi,\pi]^d}
   \frac{d^d\pos{p}}{(2\pi)^d} \frac{1}{K(\pos{p}) 
(K(\pos{p})+\lambda_V)} +
    \Sigma,
\end{equation}
where
\begin{equation}
\Sigma \equiv \frac{1}{V}\sum_{\pos{p} \in
   \Lambda_V^*}\frac{1}{K(\pos{p})+\lambda_V} -
\int_{[-\pi,\pi]^d} \frac{d^d\pos{p}}{(2\pi)^d} 
\frac{1}{K(\pos{p})+\lambda_V}.
\end{equation}
Now, for $D<4$ and $\lambda_V\to 0$ we can expand $K(\pos{p})$ in the 
integral
appearing in Eq.~(\ref{eq:app-gapeq}) and extend the integration over 
all
$\mathbb{R}^d$, obtaining
\begin{equation}
\label{appA:gap}
\beta - \beta_c \approx - \lambda_V^{D/2-1} B_1 + \Sigma,
\end{equation}
where
\begin{equation}
B_1 \equiv
\int_{\mathbb{R}^d} \frac{d^{d}\pos{p}}{(2\pi)^{d}}
     \frac{1}{K_c(\pos{p})(K_c(\pos{p})+1)} = 
   A_{p,q,\rho,\sigma},
\end{equation}
with $K_c(\pos{p}) = |\pos{p}_\bot|^{2\rho} + |\pos{p}_\||^{2 \sigma}$
and $A_{p,q,\rho,\sigma}$ defined in Eq.~(\ref{def-Apq}).
Using the Poisson summation formula for a periodic function 
$f(x) = f(x+l)$, 
\begin{equation}
   \sum_{n=0}^{l-1} f(n) = 2\pi \sum_{k=-\infty}^\infty 
    \int_0^l dx f(x) e^{2\pi i kx} ,
\label{eq:poisson}
\end{equation}
we can rewrite
\begin{equation}
\label{eq:ursigma}
\Sigma  =  \sum_{\pos{n}\neq\pos{0}} \int_{[-\pi,\pi]^d}
\frac{d^d\pos{p}}{(2\pi)^d} \frac{e^{i \pos{p}_\| \pos{n}_\| M+i
     \pos{p}_\bot \pos{n}_\bot L}}{K(\pos{p})+\lambda_V} 
\approx
\sum_{\pos{n}\neq\pos{0}} \int_{\mathbb{R}^d}
\frac{d^d\pos{p}}{(2\pi)^d} \frac{e^{i \pos{p}_\| \pos{n}_\| M+i
     \pos{p}_\bot \pos{n}_\bot L}}{K_c(\pos{p})+\lambda_V},
\end{equation}
where in the last step we have assumed $L,M\to \infty$.
Convergence is guaranteed by the oscillating phase factor.

In order to compute the asymptotic behavior of $\Sigma$ for
$\lambda_V\to 0$, $M,L\to \infty$, we introduce the Laplace
transform $f_\alpha(p)$ of $\exp(-u^\alpha)$, i.e. we define
\begin{equation}
\int^\infty_0 dp\, e^{-u p} f_\alpha(p) = e^{-u^\alpha},
\end{equation}
for $0<\alpha\le 1$. Of course, for $\alpha = 1$, $f_1(p) = \delta(p-1)$.
It is easy to show that for $0<\alpha<1$ 
\begin{equation}
f_\alpha(p) \sim p^{-(1+\alpha)}, \qquad \text{for $p \to +\infty$},
\end{equation}
and {
\begin{equation}
f_\alpha(p) \sim p^{-(2-\alpha)/[2(1-\alpha)]}
   \exp\left[-(1-\alpha) (p/\alpha)^{-\alpha/(1-\alpha)}\right],
    \qquad \text{for $p \to 0+$}.
\end{equation}
Then, we obtain
\begin{eqnarray}
\Sigma & = & \sum_{\pos{n} \neq \pos{0}}\int_{\mathbb{R}^d}
\frac{d^d\pos{p}}{(2\pi)^d} \int_0^\infty dt\,
e^{-\lambda_V t -t |\pos{p}_\bot|^{2\rho} -t |\pos{p}_\||^{2\sigma} +i 
\pos{p}_\bot \cdot \pos{n}_\bot L + i \pos{p}_\| \pos{n}_\| M}
\nonumber \\
& = & \sum_{\pos{n} \neq \pos{0}}
     \int_0^\infty d\eta d\tau\, f_\rho(\eta)f_\sigma(\tau)
     \int_{\mathbb{R}^d} \frac{d^d\pos{p}}{(2\pi)^d}
      \int_0^\infty dt\,
e^{-\lambda_V t -\eta_t |\pos{p}_\bot|^{2} -\tau_t |\pos{p}_\||^{2} +
      i \pos{p}_\bot \cdot \pos{n}_\bot L + i \pos{p}_\| \pos{n}_\| M}
\nonumber \\
& = & \frac{L^{(2-D)\rho}}{(4 \pi)^{\rho+(d-\rho D)/2}}
    \int_0^\infty d\eta d\tau\, f_\rho(\eta)f_\sigma(\tau) H_{p,q}(\eta,\tau,z,r)
   \equiv \lambda_V^{D/2-1} I_{p,q,\rho,\sigma}(z,S),
\label{sigma-as}
\end{eqnarray}
where $S \equiv M/L^{\rho/\sigma}$, $r \equiv (4\pi)^{\rho/\sigma-1} S^2$, 
$z^2 \equiv (4\pi)^{-\rho} L^{2 \rho} \lambda_V$, 
$\eta_t \equiv t^{1/\rho}\eta$, $\tau_t \equiv t^{1/\sigma}\tau$,
\begin{equation}
H_{p,q}(\eta,\tau,z,r) \equiv
\int_0^\infty dt\, \eta_t^{-p/2} \tau_t^{-q/2}  e^{-z^2 t }
\left[ B^{p}\left(\eta_t^{-1}\right) B^q\left(r\tau_t^{-1}\right) 
-1\right],
\label{def-H}
\end{equation}
and
\begin{equation}
B(s) \equiv \sum_{n \in \mathbb{Z}} e^{-\pi n^2 s}.
\end{equation}
Note the well-known property
\begin{equation}
B(s) = s^{-1/2} B(1/s),
\label{B-duality}
\end{equation}
that implies $B(s)\approx s^{-1/2}$ for $s\to 0$.

Using Eqs.~(\ref{appA:gap}) and (\ref{sigma-as}) we obtain 
Eq.~(\ref{eq:gap_equation_sconst}). Note that for an isotropic geometry
with $q=0$, $I_{p,0,\rho,0}(z,S)$ does not depend on $S$ and thus we will
simply write $I^{\rm iso}_{p,\rho}(z) \equiv I_{p,0,\rho,0}(z,S)$. 
For a geometry with $M=\infty$---we call it layer 
geometry---Eq.~(\ref{eq:gap_equation_sconst}) still holds with $S=\infty$.
In Eq.~(\ref{def-H}) it corresponds to setting $r=\infty$, i.e. 
replacing $B^q(r\tau_t^{-1})$ with 1.

In the low-temperature phase ($\beta > \beta_c$ fixed) we have
$\lambda_V\to 0$ for all $\beta$ when 
$V\rightarrow\infty$. Thus, we can use the
above-reported expressions to determine the infinite-volume behavior 
of $\xi_{\bot,V}$.
{}From Eq.~(\ref{appA:gap}) we obtain $\Sigma = \beta - \beta_c > 0$
in the infinite-volume limit. Because of the prefactor $L^{(2-D)\rho}$
appearing in the last term of Eq.~(\ref{sigma-as}), since $D>2$,
this is possible only if the integral, i.e. $H_{p,q}(\eta,\tau,z,r)$ diverges
(the $\eta$ and $\tau$ integrations are completely harmless),
which only happens for $z\to 0$. In order to compute the
asymptotic behavior of $H_{p,q}(\eta,\tau,z,r)$ for $z\to 0$, we use the
duality property (\ref{B-duality}) to rewrite
\begin{eqnarray}
H_{p,q}(\eta,\tau,z,r) &=&
\int_0^1 dt\, \eta_t^{-p/2} \tau_t^{-q/2}  e^{-z^2 t}
\left[ B^{p}\left(\eta_t^{-1}\right) B^q\left(r\tau_t^{-1}\right) 
-1\right]
\nonumber \\
&& \quad
   + \int_1^\infty dt\, r^{-q/2}  e^{-z^2 t}
\left[ B^{p}\left(\eta_t\right) B^q\left(\tau_t/r\right) -1\right]
\nonumber \\
&& \quad
   + \int_1^\infty dt\, e^{-z^2 t}
   \left( r^{-q/2} - \eta_t^{-p/2} \tau_t^{-q/2} \right).
\label{H-expr2}
\end{eqnarray}
The first two integrals are always finite while the third one gives
$H_{p,q}(\eta,\tau,z,r) \approx r^{-q/2}/z^2$.
This implies that in the infinite-volume limit
\begin{equation}
   \label{eq:lowt-scaling}
   \xi_{\bot,V} \sim S^{q/2\rho}L^{\rho D/2}.
\end{equation}

\subsection{The case $D<2$} \label{AppA.2}

For $D<2$ little changes. Now criticality is observed only for 
$\beta\rightarrow\infty$. We rewrite the gap equation as
\begin{equation}
\label{eq:app-gapeq2}
\beta =
   \int_{[-\pi,\pi]^d}
   \frac{d^d\pos{p}}{(2\pi)^d} \frac{1}{K(\pos{p})+\lambda_V} +
    \Sigma .
\end{equation}
For $D<2$ and $\lambda_V\rightarrow 0$, we can expand $K(\pos{p})$ in the 
first term and extend the integration over all $\mathbb{R}^d$, while
$\Sigma$ can be treated as before. We obtain
\begin{equation}
\beta = \lambda_V^{D/2-1} B_2 + \lambda_V^{D/2-1} I_{p,q,\rho,\sigma}(z,S),
\end{equation}
where
\begin{equation}
B_2 \equiv \int_{\mathbb{R}^d}
   \frac{d^d\pos{p}}{(2\pi)^d} \frac{1}{K_c(\pos{p})+ 1} = 
   - A_{p,q,\rho,\sigma},
\end{equation}
and $A_{p,q,\rho,\sigma}$ is defined in Eq.~(\ref{def-Apq}). 
Eq.~(\ref{eq:gap_equation_sconst2}) follows immediately.

\section{FSS of the correlation function}
\label{AppB}

Now we consider the (real-space) correlation
function $G_V(\pos{x})$:
\begin{equation}
   \label{eq:FSScorrfunc1}
   G_V(\pos{x}) = \frac{1}{V}\sum_{\pos{p} \in \Lambda^*_V}
   \frac{e^{i \pos{p}\cdot \pos{x}}}{K(\pos{p})+\lambda_V},
\end{equation}
and compute its asymptotic behavior for $L,M,|x|\to\infty$ and 
$\lambda_V\to0$.
Using the Poisson summation formula (\ref{eq:poisson}) we obtain
\begin{equation}
G_V(\pos{x})  =
  \int_0^\infty dt\, e^{-\lambda_V t}  \sum_{\pos{n}}
  \int_{[-\pi,\pi]^d} \frac{d^d\pos{p}}{(2\pi)^d}
   e^{i \pos{p}_\| (\pos{n}_\| M+\pos{x}_\|)+
      i \pos{p}_\bot (\pos{n}_\bot L+\pos{x}_\bot)
      - t K(\pos{p})}\; .
\end{equation}
Since we are interested in the limit $|x|,L,M\to\infty$
we can replace $K(\pos{p})$
with its small-$\pos{p}$ expansion and extend the integration to the 
whole space, obtaining
\begin{eqnarray}
G_V(\pos{x})
  &\approx& \frac{1}{(4\pi)^{d/2}}
   \int_0^\infty d\eta d\tau f_\rho(\eta) f_\sigma(\tau)
\nonumber \\
  && \times
   \int_0^\infty dt\, e^{-\lambda_V t}
   \tau_t^{-q/2} \eta_t^{-p/2} \sum_{\pos{n} \in \mathbb{Z}^d} 
e^{-|\pos{x}_\bot+L
   \pos{n}_\bot|^2/(4 \eta_t) - |\pos{x}_\|+M \pos{n}_\||^2/(4 \tau_t)}.
\end{eqnarray}
By rescaling $t$, we obtain finally
\begin{equation}
\begin{split}
G_V(\pos{x}) & \approx
   \frac{L^{\rho(2-D)}}{(4\pi)^{d/2}}
   \int_0^\infty d\eta d\tau f_\rho(\eta) f_\sigma(\tau)
   \int_0^\infty dt\, e^{-(\lambda_V L^{2\rho}) t} \tau_t^{-q/2} 
\eta_t^{-p/2}
\\
& \qquad \qquad \qquad
    \sum_{\pos{n} \in \mathbb{Z}^d} e^{-|(\pos{x}_\bot/L)+
   \pos{n}_\bot|^2/(4 \eta_t) - S^2 |(\pos{x}_\|/M)+ \pos{n}_\||^2/(4 
\tau_t)}
\\ & = L^{\rho(2-D)} F(\pos{x}_\bot/L, \pos{x}_\|/M, S, \lambda_L 
L^{2\rho})
\\ & = \xi_\bot^{\rho(2-D)}
   \widetilde G (\pos{x}_\bot/\xi_\bot, \pos{x}_\|/\xi_\|, S, 
\xi_\bot/L,S).
\end{split}
\label{eq:oinfty-fss-corrfunc}
\end{equation}


\section{Gap equation for $S\rightarrow\infty$} \label{AppC}

\subsection{General results} \label{AppC.1}

Now we will determine the limiting form of the gap
equation~(\ref{eq:app-gapeq}) for $L \to \infty$, $M \to \infty$,
$S \to \infty$ and arbitrary values of 
$z \equiv (4\pi)^{-\rho/2} (L/\xi_{\bot,V})^\rho$, {\em including}
$z = 0$. Indeed, we shall mainly be interested in the limit $z\to 0$
together with $S\to \infty$.

For this purpose we need to compute the asymptotic behavior of
$H_{p,q}(\eta,\tau,z,r)$ (see Eq.~(\ref{def-H})) 
for $r\to\infty$. We start from Eq.~(\ref{H-expr2}).
For what concerns the first term in this equation, we can take the limit
naively, i.e. we can replace $B(r\tau_t^{-1})$ with 1. The second term
can be rewritten in the form 
\begin{eqnarray}
&& r^{-q/2} \int_1^\infty dt\, e^{-z^2 t}
   [ B^p(\eta_t) B^q(\tau_t/r) - 1] =
   r^{-q/2} \int_1^\infty dt\, e^{-z^2 t}
   [ B^p(\eta_t) -1] B^q(\tau_t/r)
\nonumber \\
&& \qquad + r^{-q/2} \int_1^\infty dt\, e^{-z^2 t}  B^q(\tau_t/r) -
             r^{-q/2} \int_1^\infty dt\, e^{-z^2 t}.
\end{eqnarray}
In the first term we can simply replace $B(\tau_t/r)$ with its
asymptotic behavior for $\tau_t/r$ small, i.e. $(\tau_t/r)^{-1/2}$, 
while some additional
manipulations are need for the second one. Rescaling
$t = s r^\sigma$, we rewrite it as
\begin{eqnarray}
&& r^{-q/2} \int_1^\infty dt\, e^{-z^2 t}  B^q(\tau_t/r) =
    r^{\sigma-q/2} \int_1^\infty ds\, e^{-z^2 r^\sigma s} [ B^q(\tau_s) 
- 1]
\nonumber  \\
&& \qquad + r^{\sigma-q/2} \int_{r^{-\sigma}}^1 ds\, e^{-z^2 r^\sigma s}
    \tau_s^{-q/2} [ B^q(\tau_s^{-1}) - 1] +
    r^{\sigma-q/2} {e^{-z^2 r^\sigma} \over z^2 r^\sigma}
\nonumber  \\
&& \qquad + r^{\sigma-q/2} \tau^{-q/2}
    \int_{r^{-\sigma}}^1 ds\, e^{-z^2 r^\sigma s} s^{-q/2\sigma}.
\end{eqnarray}
In the second term, one can easily convince himself that it is safe to
replace the lower integration limit $r^{-\sigma}$ with 0.
Collecting everything together, we obtain
\begin{eqnarray}
H_{p,q}(\eta,\tau,z,r) &=&
\tau^{-q/2}  \int_0^1 dt\, e^{-z^2 t} t^{-q/2\sigma} \eta_t^{-p/2} 
\left[
  B^{p}(\eta_t^{-1}) -1\right]
\nonumber \\ && \quad
+ \tau^{-q/2}  \int_1^\infty dt\, e^{-z^2 t} t^{-q/2\sigma} \left[
  B^{p}(\eta_t) -1\right]
\nonumber \\ && \quad
+   r^{\sigma-q/2} \int_1^\infty dt\, e^{-z^2 r^\sigma t }
  \left[B^q (\tau_t) -1\right ]
\nonumber \\ && \quad
+   r^{\sigma-q/2} \int_{0}^1 dt\, e^{-z^2 r^\sigma t }
      \tau_t^{-q/2} \left[B^q\left( \tau_t^{-1} \right)-1\right]
\nonumber \\ && \quad
+ r^{\sigma-q/2} \frac{e^{-z^2 r^\sigma}}{z^2 r^\sigma} +
  \tau^{-q/2} \int_1^{r^\sigma} dt\, e^{-z^2 t } t^{-q/2\sigma}
\nonumber \\ && \quad
  -\tau^{-q/2} \eta^{-p/2} \int_1^\infty dt\, t^{-D/2} e^{-z^2 t} .
\end{eqnarray}
Then, for $D>2$, we can write the gap equation as
\begin{eqnarray}
&& (4\pi)^{\rho(2-D)/2+ d/2} (\beta-\beta_c)L^{\rho(D-2)} =
-(4\pi)^{d/2} A_{p,q,\rho,\sigma} z^{D-2} 
\nonumber \\ 
&& \qquad + r^{\sigma-q/2}{\cal J}_1(z^2 r^\sigma ) +  {\cal J}_2(z^2) +
C_{\sigma,q} \int_1^{r^\sigma}\! dt\, e^{-z^2 t} t^{-q/2\sigma},
\label{gapeqlim2}
\end{eqnarray}
where 
\begin{eqnarray}
{\cal J}_1(x) &\equiv& \frac{e^{-x}}{x} + \mathcal{G}_{\sigma,q}^{(1)}(x), \\
{\cal J}_2(x) &\equiv&
  -C_{\sigma,q} C_{\rho,p} \int_1^\infty dt\, t^{-D/2} e^{-x t}
       + \mathcal{G}_{\rho,p,\sigma,q}^{(2)}(x),
\end{eqnarray}
and 
\begin{eqnarray}
C_{\alpha,\beta} &\equiv& \int_0^\infty dp\, p^{-\beta/2} f_\alpha(p) =
\frac{1}{\alpha}\frac{\Gamma(\beta/2\alpha)}{\Gamma(\beta/2)},
\nonumber \\
\mathcal{G}_{\sigma,q}^{(1)}(x) &\equiv&
       \int_0^\infty d\tau\, f_\sigma(\tau) G_{\sigma,q}^{(1)}(\tau,x),
\nonumber \\
G_{\sigma,q}^{(1)}(\tau,x) &\equiv&
  \int_{1}^\infty dt\, e^{-x t }
\left[   B^q\left( \tau_t \right) -1\right]   +
\int_{0}^1 dt\, e^{-x t } \tau_t^{-q/2}
      \left[B^q\left( \tau_t^{-1} \right)-1\right]
\nonumber \\  &=&
\int_{0}^\infty dt\, e^{-x t } \tau_t^{-q/2}
      \left[B^q\left( \tau_t^{-1} \right)-1\right] +
  \tau^{-q/2} \int_1^\infty dt\, e^{-xt} t^{-q/2\sigma}
  - \frac{e^{-x}}{x} ,
\nonumber \\
\mathcal{G}_{\rho,p,\sigma,q}^{(2)}(x) &\equiv&
   C_{\sigma,q} \int_0^\infty d\eta\, f_\rho(\eta) 
   G_{\rho,p,\sigma,q}^{(2)}(\eta,x),
\\
G_{\rho,p,\sigma,q}^{(2)}(\eta,x) &\equiv&
       \int_{1}^\infty dt\, e^{-x t } t^{-q/2\sigma}
\left[   B^p\left( \eta_t \right) -1\right]   +
\int_{0}^1 dt\, e^{-x t } t^{-q/2\sigma} \eta_t^{-p/2} 
      \left[B^p\left( \eta_t^{-1} \right)-1\right] .
\nonumber 
\end{eqnarray}
Note that $\mathcal{G}_{\sigma,q}^{(1)}(x)$, 
$\mathcal{G}_{\rho,p,\sigma,q}^{(2)}(x)$, and ${\cal J}_2(x)$
are finite for $x\rightarrow 0$ and the relation
\begin{eqnarray}
\tau^{-q/2} G_{\rho,p,\sigma,q}^{(2)}(\eta,z^2) &=& 
 H_{p,q}(\eta,\tau,z,\infty) + 
 \tau^{-q/2} \eta^{-p/2} \int_1^\infty dt\, e^{-z^2 t} t^{-D/2} 
\nonumber \\
  && - \tau^{-q/2} \int_1^\infty dt\, e^{-z^2 t} t^{-q/2\sigma}.
\label{relation-G2-H}
\end{eqnarray}
We will be interested in computing the behavior of 
${\cal J}_1(x)$ for $x\to\infty$. If $0<\sigma\le 1$, we rewrite
\begin{equation}
{\cal J}_1(x) = \int_0^\infty d\tau\, f_\sigma(\tau) 
\int_{0}^\infty dt\, e^{-x t } \tau_t^{-q/2}
      \left[B^q\left( \tau_t^{-1} \right)-1\right] +
  C_{\sigma,q} \int_1^\infty dt\, e^{-xt} t^{-q/2\sigma} .
\end{equation}
By performing an integration by parts it is easy to see that the last
term vanishes as $e^{-x}/x$. The asymptotic behavior of the first term 
depends on $\sigma$. 
For $\sigma = 1$ we can rewrite it
\begin{eqnarray}
\int_{0}^\infty dt\, e^{-x t } t^{-q/2}
      \left[B^q\left( t^{-1} \right)-1\right]  &\approx&
      2q \int_{0}^\infty dt\, t^{-q/2}
      e^{-xt - \pi t^{-1}}
\nonumber       \\ 
& = & 2q x^{q/4-1/2} \int_{0}^{\infty} dt\, t^{-q/2} e^{-\sqrt{x}(t+\pi/t)} 
\nonumber  \\ 
&\approx & 2q \left(\frac{\pi}{x}\right)^{3/4-q/4} e^{-2\sqrt{\pi x}}\; .
\end{eqnarray}
For $0 < \sigma < 1$, more work is needed. First,
we replace $\tau$ with $s = \tau t^{1/\sigma}\equiv \tau_t$ obtaining
\begin{equation}
\int_0^\infty dt\, t^{-1/\sigma} e^{-x t }
\int_{0}^\infty ds\, s^{-q/2} f_\sigma(s t^{-1/\sigma})
   \left[B^q\left( s^{-1} \right)-1\right] .
\label{eq75}
\end{equation}
In the limit of large $x$, only the small-$t$ region contributes
to the integral (because of the exponential factor $e^{-x t}$) and
the function $f_\sigma(s t^{-1/\sigma})$ can be replaced with its
small-$t$ behavior (this is safe because the integral over $s$
is convergent near $s=0$), i.e. we can replace $f_\sigma(x)$ with  
$x^{-1-\sigma}$. Thus, the integral (\ref{eq75}) becomes
\begin{eqnarray}
&& 
\int_0^\infty dt \, t\, e^{-x t} \int_{0}^\infty ds\,
s^{-1-\sigma-q/2} \left[B^q\left( s^{-1} \right)-1\right] 
\nonumber \\
& = & \frac{1}{x^2} \int_{0}^\infty ds\,
s^{-1-\sigma-q/2} \left[B^q\left( s^{-1} \right)-1\right],
\end{eqnarray}
and it is easy to check that for $\sigma>0$, $q>0$ the second integral
is convergent.  Therefore, as $x \to \infty$, we have that 
${\cal J}_1(x) \sim x^{-2}$ for any $0 < \sigma < 1$ and 
${\cal J}_1(x) \sim x^p \exp (- C\sqrt{x})$ for $\sigma = 1$.

From Eq.~(\ref{gapeqlim2}) we see that there are two interesting scaling
limits: (a) $r\to \infty$ at fixed $z$; 
(b) $r\to \infty$ at fixed $z^2 r^\sigma$. As we will show below these two 
limits correspond to the two different cases we discussed in Sec. \ref{sec4}.
They are descussed in detailed below.

\subsection{Limit (a)} \label{AppC.2}

In this case we should consider the limit $r\to \infty$ at fixed $z$ 
and we should therefore reobtain Eq.~(\ref{eq:gap_equation_sconst}) 
with $S=\infty$. This follows immediately from the fact that 
$r^{\sigma-q/2} {\cal J}_1 (z^2 r^\sigma) \to 0$, cf. Sec.~\ref{AppC.1}, 
and from 
\begin{eqnarray}
{\cal J}_2(z^2) + C_{\sigma,q}\int_1^\infty dt\, e^{-z^2t} t^{-q/2\sigma} = 
\int_0^\infty d\eta d\tau\, f_\rho(\eta)f_\sigma(\tau) 
   H_{p,q}(\eta,\tau,z,\infty),
\end{eqnarray}
which is a direct consequence of Eq.~(\ref{relation-G2-H}).

\subsection{Limit (b)} \label{AppC.3}

Now we discuss the limit $z^2\rightarrow 0$ and 
$r\rightarrow \infty$ at constant $z^2r^\sigma$. Note that the last
term in Eq.~(\ref{gapeqlim2}) behaves differently, depending on 
whether $q/2\sigma$ is larger or smaller than $1$. For $q/\sigma>2$,
integrating by parts, we obtain
\begin{eqnarray}
\int_1^{r^\sigma} dt e^{-z^2t}t^{-q/2\sigma} &= &
\frac{2\sigma}{q-2\sigma}e^{-z^2} \nonumber \\
&& -\frac{2\sigma}{q-2\sigma} r^{\sigma-q/2}\left[e^{-z^2r^\sigma} +
z^2r^\sigma\int_{r^{-\sigma}}^1 dt e^{-z^2t}t^{1-q/2\sigma} \right] .
\label{eq:max1}
\end{eqnarray}
Then, note that for $D<4$ we can extend the remaining integration down to 
zero. Moreover,  we can approximate $e^{-z^2}$ with $1$ since $z^2$ is 
subleading with respect to $r^{\sigma-q/2}$: indeed
$z^2\sim r^{-\sigma} \sim r^{\sigma-q/2}\times r^{q/2-2\sigma}$ and
$q<4\sigma$ for $D<4$.
Therefore, if we define
\begin{eqnarray}
K &\equiv & {\mathcal J}_2(0) + \frac{2\sigma}{q-2\sigma} C_{\sigma,q}, 
\label{def-K} \\
\widehat{\mathcal J}_1(x) &\equiv & {\mathcal J}_1(x) - 
\frac{2\sigma}{q-2\sigma} C_{\sigma,q} \left[e^{-x} + x \int_0^1 dt
e^{-xt} t^{1-q/2\sigma}\right]  \nonumber \\
&=& \int_0^\infty d\tau dt f_\sigma(\tau) e^{-xt}\tau_t^{-q/2}
[B^q(\tau_t^{-1})-1] + C_{\sigma,q}\Gamma(1-q/2\sigma) 
x^{(q-2\sigma)/2\sigma} ,
\label{defhatJ}
\end{eqnarray}
we obtain
\begin{equation}
(4\pi)^{\rho+(d-\rho D)/2} (\beta-\beta_c) 
L^{\rho(D-2)} = r^{\sigma-q/2} \widehat{\mathcal{J}}_1(z^2 r^\sigma) + K,
\label{eq:max0}
\end{equation}
where we have also discarded the term proportional to 
$z^{D-2} \sim (z r^{\sigma/2})^{D-2} r^{\sigma-q/2} r^{-\sigma p/2\rho }$,
which is subleading with respect to $r^{\sigma-q/2}$.

For $q/\sigma < 2$, we write
\begin{eqnarray}
\int_1^{r^\sigma} e^{-z^2 t} t^{-q/2\sigma} & = &
  r^{\sigma-q/2} \int^1_{r^{-\sigma}} e^{-z^2 r^\sigma t} t^{-q/2\sigma} \,dt 
\nonumber \\
    & = &  r^{\sigma-q/2} \int^1_{0} e^{-z^2 r^\sigma t} t^{-q/2\sigma} \,dt 
-  \int^1_{0} e^{-z^2 t} t^{-q/2\sigma} \,dt.
\label{eq:caso3}
\end{eqnarray}
It is then easy to see that we reobtain Eq.~(\ref{eq:max0}). In this case
$K$ is subleading and can be neglected in the scaling limit.

In this derivation we have implicitly assumed that it is possible to
take the limit $z\to 0$, $r\to \infty$, with $z^2 r^\sigma$ fixed. 
However, we now show that this is not the case for $\beta < \beta_c$. 
Indeed, the function $\widehat{\cal J}_1(z)$ is positive and 
thus Eq.~(\ref{eq:max0})
can only be valid if $\beta>\beta_c$. If this condition is not satisfied, 
the only possibility is that $(\beta - \beta_c) L^{\rho(D-2)}$ vanishes in the 
scaling limit as well as the right-hand side, which in turn implies 
$z^2 r^\sigma \to \infty$. This means $\lambda_V M^{2\sigma} \to \infty$, 
or $\chi/M^{2\sigma} \to 0$.

Finally, we rewrite the gap equation in terms of the constant
$A_{q,\sigma}\equiv A_{0,q,\rho,\sigma}$ (it is easy to check that 
it is $\rho$-independent) and of the integral $I_{q,\sigma}^{\rm iso}(z)$
that appear in the gap equation of an isotropic $q$-dimensional 
system. It is indeed easy to show that 
\begin{equation}
\widehat{\cal J}_1(z^2) = (4\pi)^{q/2} z^{-2+q/\sigma} 
   \left[I_{q,\sigma}^{\rm iso}(z) - A_{q,\sigma}\right].
\end{equation}

\end{document}